\documentclass[aps,prd,twocolumn,superscriptaddress,showpacs,amsmath,amssymb]{revtex4}
\usepackage{graphicx,epsf}
\usepackage[]{latexsym}
\newcommand{\be}{\begin{eqnarray}}
\newcommand{\ee}{\end{eqnarray}}
\newcommand{\pa}{\partial_a}
\newcommand{\pt}{\partial_t}
\newcommand{\bra}[1]{\mbox{$\langle #1 \mid$}}

\newcommand{\ket}[1]{\mbox{$\mid \!#1\rangle$}}

\newcommand{\expec}[1]{\mbox{$\langle #1\rangle$}}

\newcommand{\real}{\mbox{{\rm I\hspace{-2truemm} R}}}

\newcommand{\re}{\real{\rm e}}

\renewcommand{\a}{\hat a}
\newcommand{\ac}{\hat a^{\dagger}}
\renewcommand{\b}{\hat b}
\newcommand{\bc}{\hat b^\dagger}
\renewcommand{\d}{\mbox{{\rm d}}}

\begin{document}
\title{Planck scale inflationary spectra from quantum gravity}
\author{Gian~Luigi~Alberghi}
\email{alberghi@bo.infn.it}
\affiliation{Dipartimento di Fisica, Universit\`a di Bologna,
and I.N.F.N., Sezione di Bologna, via~Irnerio~46, 40126~Bologna, Italy}
\affiliation{Department of Astronomy, University of Bologna,
via~Ranzani~1, 40127~Bologna, Italy.}
\author{Roberto~Casadio}
\email{casadio@bo.infn.it}
\affiliation{Dipartimento di Fisica, Universit\`a di Bologna,
and I.N.F.N., Sezione di Bologna, via~Irnerio~46, 40126~Bologna, Italy}
\author{Alessandro~Tronconi}
\email{tronconi@bo.infn.it}
\affiliation{Dipartimento di Fisica, Universit\`a di Bologna,
and I.N.F.N., Sezione di Bologna, via~Irnerio~46, 40126~Bologna, Italy}
\begin{abstract}
We derive the semiclassical evolution of massless minimally coupled scalar
matter in the de~Sitter space-time from the Born-Oppenheimer reduction of
the Wheeler-DeWitt equation.
We show that the dynamics of trans-Planckian modes can be cast in the form
of an effective modified dispersion relation and that
high energy corrections in the power spectrum of the
cosmic microwave background radiation produced during inflation
remain very small if the initial state is the Bunch-Davies vacuum.
\end{abstract}
\maketitle
\section{Introduction}
It is commonly accepted that inflation~\cite{infla} is a
viable paradigm for the early Universe which solves some
of the problems of the standard big-bang scenario and allows one
to make testable predictions about the spectrum of the cosmic
microwave background radiation~(CMBR).
However, it has also been realized that inflation provides a
window towards trans-Planckian physics~\cite{branden} as it
magnifies all quantum fluctuations and red-shifts originally
trans-Planckian frequencies down to the range of low energy
physics currently observed.
This has raised two related issues:
a) how to describe the original quantum fluctuations in such
a high energy regime, given that there is currently no
universally accepted theory of quantum gravity,
and 
b) whether the effect of now red-shifted trans-Planckian frequencies
can be observed with the precision of present and future experiments.
\par
One can take the pragmatic view of the renormalization group
approach to high energy physics and use quantum field theory on
a classical metric background to describe quantum fluctuations
after their frequencies have been red-shifted below the scale
of quantum gravity.
However, the higher energy dynamics cannot be {\em a priori\/}
neglected as it could enter (at least) in the form of initial
conditions for the quantum state of the field theory~\cite{ulf,bozza}
and, in turn, affect the CMBR spectrum (to an extent which is
currently being debated;
see, e.g.,~Refs.~\cite{ulf,bozza,tp,kaloper,act}).
\par
As a consequence of the above observations, one expects that the
dispersion relations
of matter fields change on approaching the Planckian regime,
so that the frequency
$\omega$ of massless modes will depend on their wavenumber $k$
according to an expression of the general form
\be
\omega=\frac{k}{a}\,\left[1
+f\left(k,\ell_{\rm p}/a,\dot a,\ldots\right)
\right]
\ ,
\label{mod_disp}
\ee
in which $\ell_{\rm p}$ is the Planck length and $a=a(t)$
the scale factor of the flat Robertson-Walker metric~\cite{mtw},
\be
\d s^2=
-N^2\,\d t^2+a^2\left(\d \vec x\right)^2
\ .
\label{rw}
\ee
Most attempts have tested the effects of functions $f$
chosen {\em ad hoc\/} in Eq.~(\ref{mod_disp})~\cite{branden}.
We shall instead {\em derive\/} the dispersion relation for
a minimally coupled massless scalar field from the
action principle previously employed to study the semiclassical
dynamics of non-minimally coupled scalar fields in Ref.~\cite{acg}.
\par
In that paper, the principle of time-reparameterization
invariance was lifted to a quantum symmetry.
We then obtained an Hamiltonian constraint from which the
Born-Oppenheimer (BO) reduction~\cite{bv} allowed to properly
and unambiguously recover the semiclassical limit of quantum
field theory on a curved background starting from the
Wheeler-DeWitt (WDW) equation~\cite{dewitt}.
This required that certain ``quantum fluctuation'' terms be negligible
with respect to the usual matter contributions.
Precisely such terms in the matter equation will now be studied
and their effect on the power spectrum derived for the simplest
model of inflation.
\par
In the next Section, we briefly introduce the model and its classical
and Wheeler-DeWitt dynamics in the BO decomposition
(for more details, see Ref.~\cite{acg}).
In Section~\ref{s_mat_eq}, we then focus on the quantum equation
for matter which we solve perturbatively in order to determine a
dispersion relation and the corresponding CMBR spectrum.
Finally, we comment on our findings in Section~\ref{s_conc}.
We shall use units with $c=\hbar=1$.
\section{Scalar field in de~Sitter space}
As the simplest model of inflation~\footnote{More general
cases will be studied in Ref.~\cite{toappear}.},
we shall consider the de~Sitter metric~(\ref{rw}) with
\be
a=a_0\,\exp({\mathcal H}\,t)
\ee
and cosmological constant
$\Lambda=3\,{\mathcal H}^2$.
The massless minimally coupled scalar field $\phi$ will be
treated as a perturbation with respect to $\Lambda$
in the interval $t_{\rm i}\le t\le t_{\rm f}$,
where the relevant modes $k$ were first produced at
$t=t_{\rm i}$ and inflation ends at $t=t_{\rm f}$.
Upon varying the corresponding action
($\dot f\equiv\partial_t f$),
\be
S&\!\!=\!\!&
-\frac{1}{2\,\ell_{\rm p}^2}
\int_{t_{\rm i}}^{t_{\rm f}}
\!\!\!\!
a^3 N\,\d t
\left(\frac{\dot a^2}{N^2 a^2}+\frac{\Lambda}{3}\right)
\nonumber
\\
&&
\ \ \
+
{\frac{1}{2}
\int_{t_{\rm i}}^{t_{\rm f}}
\!\!\!\!
a^3 N\,\d t}
\int
\d^3 x\left[
\frac{\dot\phi^2}{N^2}
-\frac{(\vec\partial\phi)^2}{a^2}
\right]
\ ,
\label{Sp}
\ee
one obtains the Euler-Lagrange equations of motion
in the form of the Hamiltonian constraint
\begin{subequations}
\be
\delta_N S\equiv -H=0
\ ,
\label{H=0}
\ee
and the dynamical equations
\be
\delta_a S=\delta_\phi S=0
\ .
\label{nonH}
\ee
\end{subequations}
The former allows one to work in the gauge $N(t)=1$,
provided the initial conditions are such that $H(t_{\rm i})=0$,
and Eqs.~(\ref{nonH}) will then evolve the initial data
consistently, since the Hamiltonian constraint~(\ref{H=0})
is then preserved,
\be
\dot H=-\dot a\,\delta_a S-\dot\phi\,\delta_\phi S=0
\ .
\ee
However, since
\be
H(t)=\delta_\phi S=0
\quad 
\Rightarrow
\quad
\delta_a S=0
\ ,
\label{dHH}
\ee
for our purposes it is more convenient to impose the Hamiltonian
constraint together with the Klein-Gordon equation at all
times.
\subsection{WDW~equation}
At the quantum level, Eq.~(\ref{H=0}) becomes the WDW constraint
which reads
\be
H\,\ket{\Psi}=
\left(\hat H_G+\sum\nolimits_k \hat H_k\right)
\ket{\Psi}=0
\ ,
\label{iwdw}
\ee
where the gravitational Hamiltonian is~\cite{order}
\be
\hat H_G=\frac{1}{2}
\left(\ell_{\rm p}^2\,\frac{\pa^2}{a}
+\frac{a^3\,\Lambda}{3\,\ell_{\rm p}^2}
\right)
\ ,
\ee
and the matter Hamiltonian is given by the sum of
the contributions for each mode $\phi_k$ of the scalar field,
\be
\hat H_k=\frac{1}{2}
\left(
\frac{\hat\pi_k^2}{\hat a^3}
+\hat a\,k^2\,\hat \phi_k^2\right)
\ ,
\ee
with $[\hat \phi_k,\hat\pi_{k'}]=i\,\delta_{k,k'}$.
This equation is rather involved, but the following treatment
allows us to obtain more tractable equations.
\subsection{BO~decomposition}
We can decompose the wavefunction of the Universe
according to the BO~prescription~\cite{bv} as
\be
\ket{\Psi(a,\phi)}=\psi(a)\,\ket{\chi(a,\phi)}
\equiv
\psi\,\prod\nolimits_k\ket{\chi_k}
\ ,
\label{BOinh}
\ee
where $\ket{\chi_k}$ is the wavefunction for the mode $k$.
Eq.~(\ref{iwdw}) can then be shown~\cite{toappear} to be equivalent
to a system of decoupled Schwinger-Tomonaga equations
(one for each mode) and a Friedmann-like equation expressing
the back-reaction of matter on the expansion of the Universe.
In fact, starting from the ansatz~(\ref{BOinh}) one can reduce
Eq.~(\ref{iwdw}) to the two coupled equations
\begin{subequations}
\be
&&
\!\!\!\!\!\!\!\!\!\!\!\!\!\!\!\!\!\!
\left(\frac{\ell_{\rm p}^2}{2}\,D_{_-}^2
+\frac{a^4\,\Lambda}{6\,\ell_{\rm p}^2}
+a\,\expec{\hat H_\phi}
\right)\psi
=
-\frac{\ell_{\rm p}^2}{2}\,\expec{D_{_+}^2}\,\psi
\label{Fried1}
\\
&&
\!\!\!\!\!\!\!\!\!\!\!\!\!\!\!\!\!\!
\frac{\ell_{\rm p}^2}{a}
\left(D_{_-}\psi\right)
D_{_+}\ket{\chi}
+\psi\left(\hat H_\phi-\expec{\hat H_\phi}\right)
\ket{\chi}
\nonumber
\\
&&
\!\!\!\!\!\!\!
=-\frac{\ell_{\rm p}^2}{2\,a}\,\psi
\left(D_{_+}^2-\expec{D_{_+}^2}\right)
\ket{\chi}
\ ,
\label{mateq}
\ee
\end{subequations}
where $\expec{\hat O}\equiv \bra{\chi}\hat O\ket{\chi}$
and the action of the ``covariant'' derivatives
$D_{_\pm}\equiv \pa\pm\expec{\pa}$ on $\psi$ and $\ket{\chi}$
reduces to that of $\pa$ after rescaling by a geometric
phase~\cite{acg},
\begin{subequations}
\be
\ket{\chi}
&\to&
\ket{\tilde \chi}\equiv
{\rm e}^{+i\,\int^a\langle{\partial_{a'}}\rangle\,{\rm d} a'}
\ket{\chi}
\\
\psi
&\to&
\tilde\psi\equiv
{\rm e}^{-i\,\int^a\langle{\partial_{a'}}\rangle\,{\rm d} a'}
\psi
\ .
\ee
\end{subequations}
\par
Eq.~(\ref{mateq}) is still rather involved
since it contains the full matter wavefunction.
Upon projecting onto $\prod_{n\neq k}\bra{\tilde\chi_n}$,
one can obtain independent equations for each mode,
\begin{subequations}
\be
&&
\frac{\ell_{\rm p}^2}{a}
\left(\frac{\pa\tilde\psi}{\tilde\psi}\right)
\pa\ket{\tilde\chi_k}
+\left(\hat H_k-\expec{\hat H_k}_k\right)\ket{\tilde\chi_k}
\nonumber
\\
&&
\ \
=-\frac{\ell_{\rm p}^2}{2\,a}
\left(\pa^2-\expec{\pa^2}_k\right)\ket{\tilde\chi_k}
\ ,
\label{kmateq}
\ee 
where $\expec{\hat O}_k\equiv\bra{\tilde\chi_k}\hat O\ket{\tilde\chi_k}$
and the terms in the right hand side represent the kind of
``quantum gravitational fluctuations'' already mentioned in the Introduction.
Finally, one can also write the gravitational Eq.~(\ref{Fried1}) in terms
of the solutions to the matter Eq.~(\ref{kmateq}) as
\be
\!\!\!\!\!\!\!
\left(\frac{\ell_{\rm p}^2}{2\,a}\,\pa^2
+\frac{a^3\,\Lambda}{6\,\ell_{\rm p}^2}
+\sum_k\expec{\hat H_k}_k\right)
\tilde\psi
=
-\frac{\ell_{\rm p}^2}{2\,a}\,
\sum_k\expec{\pa^2}_k
\tilde\psi
.
\label{Geq}
\ee
\end{subequations}
Again, ``quantum gravitational fluctuations'' appear in the right hand side which
will affect the way matter back-reacts on the evolution of the scale factor.
However, we shall not consider the latter equation any further in the present
paper.
\section{Matter equation}
\label{s_mat_eq}
Our aim is to solve Eq.~(\ref{kmateq}) for modes with
wavelengths $a/k\lesssim \ell_{\rm p}\ll \mathcal{H}^{-1}$ at the
time $t\sim t_{\rm i}$ and evolve them to a later time $t=t_{\rm f}$
when $a/k\gg\mathcal{H}^{-1}$ in order to determine the power spectrum.
The choice of the state $\ket{\rm vac}$ is very important and
will be discussed later.
First we take the semiclassical limit for gravity which
allows us to introduce the time as~\cite{bv,tronconi}
\be
\ell_{\rm p}^2
\left(\pa\ln\tilde\psi\right)\pa
\sim i\,a\,\pt
\ .
\ee
In the infrared sector, $a/k\gg\ell_{\rm p}$,
the right~hand~side (RHS) of Eq.~(\ref{kmateq})
can be discarded, since
\be  
{\rm RHS}
\sim
(\ell_{\rm p}^2/a)\,\pa^2\ket{\tilde\chi_k}
\sim
\ell_{\rm p}^2\,({k^2}/{a^3})\ket{\tilde\chi_k}
\ee
is suppressed by a factor of order $k\,\ell_{\rm p}^2/a^2\ll 1$
with respect to $\expec{\hat H_k}\sim k/a$.
Eq.~(\ref{kmateq}) therefore coincides with the Schwinger-Tomonaga equation
for $\ket{\chi_k}$~\cite{bv,acg,tronconi}.
For the same reason, however, one expects that the RHS of Eq.~(\ref{kmateq})
significantly affects the scalar field dynamics in the ultraviolet sector
$a/k\lesssim\ell_{\rm p}$.
If we define
\be
\ket{\chi_{\rm s}}=
{\rm e}^{-i\,\int^t \langle{\hat H_k}\rangle_k\,{\rm d} t'}
\ket{\tilde\chi_k}
\ ,
\label{chis}
\ee
Eq.~(\ref{kmateq}) then becomes
\be
\left(1-\frac{3\,i\,\delta^2}{2\,a^3\mathcal{H}^3}\right)
\left(i\,\pt-\hat H_k\right)\ket{\chi_{\rm s}}
=
\frac{\delta^2}{2\,a^3\mathcal{H}^3}
\,\hat\Delta\,\ket{\chi_{\rm s}}
\ ,
\label{keqtot0}
\ee
where $\delta\equiv\mathcal{H}\,\ell_{\rm p}$ and
$\hat\Delta=\sum_{i=1}^3
\left(\hat O_i -\bra{\chi_{\rm s}}\,\hat O_i\,\ket{\chi_{\rm s}}\right)$
with
\begin{subequations}
\be
&&
\hat O_1
=
2\,{\mathcal{H}}^{-1}\,\bra{\chi_{\rm s}}\,\hat H_k\,\ket{\chi_{\rm s}}
\,i\,\pt
\\
&&
\hat O_2
=
3\,i\,\hat H_k
\\
&&
\hat O_3
=
{\mathcal{H}}^{-1}\,\pt^2
\ .
\ee
\end{subequations}
\par
Eq.~(\ref{keqtot0}) is not linear and the superposition principle no longer
holds.
However, we note that in a (semi)classical Universe the dimensionless parameter
$\delta^2\ll 1$, and we can therefore identify two regimes:
{I)}
in the very early stages $a^3\,\mathcal{H}^3\lesssim\delta^2$
matter evolves according to
\begin{subequations}
\be
\left(i\,\pt-\hat H_k\right)\ket{\chi_{\rm s}}
\simeq
\frac{i}{3}
\sum\nolimits_{i}
\hat\Delta_i\,\ket{\chi_{\rm s}}
\equiv
\hat W_{\rm I}\,\ket{\chi_{\rm s}}
\ ;
\label{keqtot1}
\ee
{II)}
after $a$ has become sufficiently large ($a\gg \delta^{2/3}\mathcal{H}^{-1}$),
Eq.~(\ref{keqtot0}) can be expanded to leading order in $\delta$ as
\be
\!\!\!\!
\left(
i\,\pt-\hat H_k
\right)
\ket{\chi_{\rm s}}
\simeq
\frac{\delta^2}{2\,a^3\mathcal{H}^3}
\sum_{i=1}^3\hat\Delta_i\,
\ket{\chi_{\rm s}}
\equiv
\delta^2
\hat W_{\rm II}\,\ket{\chi_{\rm s}}
\ .
\label{keqtot}
\ee
\end{subequations}
These expressions represent our main qualitative result:
Eq.~(\ref{keqtot}) shows that, at late stages of the cosmological
evolution, corrections coming from $\hat W_{\rm II}$ are of order
$\delta^2\ll 1$ and thus very small;
further, although the RHS of Eq.~(\ref{kmateq}) seems to produce
corrections of order $\delta^0\sim 1$ in the very early stages,
we shall see that the effect of $\hat W_{\rm I}$ is actually
negligible in Eq.~(\ref{keqtot1}).
\subsection{Perturbative analysis}
On neglecting $\hat W$ ($\equiv \hat W_{\rm I}$ or $\hat W_{\rm II}$),
the matter equations~(\ref{keqtot1}) and~(\ref{keqtot}) simply read
\be
i\,\pt\ket{\chi_{\rm s}}=
\hat H_k\ket{\chi_{\rm s}}
=\frac{k}{a}\,\left(\ac\,\a+\frac{1}{2}\right)\ket{\chi_{\rm s}}
\ ,
\label{zerosch}
\ee
where 
\be
\a\equiv\sqrt{\frac{k\,a^2}{2}}\left(\hat\phi_k
+i\,\frac{\hat \pi_k}{k\,a^2}\right)
\ ,
\ee
and $\ac$ are the usual annihilation and creation operators
such that $[\a,\ac]=1$ and can be used to evaluate the energy
of the matter state.
\par
Because of the time dependent $a=a(t)$, Hamiltonian eigenstates
defined by $\ac\,\a\,\ket{n_E}=n\,\ket{n_E}$ do not satisfy
Eq.~(\ref{zerosch}).
A basis of exact solutions of the time-dependent problem
is instead given by eigenstates of the invariant number
operator $\bc\,\b\,\ket{n}=n\,\ket{n}$, where~\footnote{For the general theory
of invariant operators in quantum mechanics see
Refs.~\cite{lewis}, and for its application to cosmological models see, {\em e.g.},
Ref.~\cite{acg,chaotic}.}
\be
\b\equiv\frac{1}{\sqrt{2}}\,
\left[\frac{\hat\phi_k}{\rho}
+i\,\left(\rho\,\hat \pi_k-a^3\,\dot\rho\,\hat\phi_k\right)
\right]
\ ,
\ee
and $\rho=\rho(t)$ must satisfy
\be
\ddot\rho+3\,{\mathcal H}\,\dot\rho+\frac{k^2}{a^2}\,\rho
=\frac{1}{a^6\,\rho^3}
\ .
\label{pinney}
\ee
One then has $[\b,\bc]=1$ and exact solutions are given by superpositions
of the base vectors
\be
\ket{n}=\frac{{\rm e}^{i\,n\,\Theta}}{n!}(\bc)^n\ket{0}
\ ,
\ee
with $\b\,\ket{0}=0$ and the phase
\be
\Theta(t)\equiv\int_0^t \frac{\d t'}{a^{3}(t')\,\rho^{2}(t')}
\ .
\ee
Note that the invariant operators $\b$ and $\bc$ (as well as
the states $\ket{n}$) are a mathematical tool to determine the solutions
and do not in general have a physical meaning.
They are however related to $\a$ and $\ac$ by the
Bogoliubov transformation~\cite{birrell}
\be
\a
&\!\!=\!\!&
B^*\,\b+A^*\,\bc
\nonumber
\\
&\!\!\equiv\!\!&
\frac{1}{2}
\left[a\,\sqrt{k}\,\rho+\frac{1}{a\,\sqrt{k}\,\rho}
+i\,\frac{a^2\dot\rho}{\sqrt{k}}\right]
\b
\nonumber\\
&&
+
\frac{1}{2}
\left[a\,\sqrt{k}\,\rho-\frac{1}{a\,\sqrt{k}\,\rho}
+i\,\frac{a^2\dot\rho}{\sqrt{k}}\right]
\bc
\ ,
\label{bogrot}
\ee 
which also relates Hamiltonian eigenstates to exact solutions.
In terms of such operators, one finds 
\begin{subequations}
\be
\hat W\,\ket{n}
=
\left[
\alpha\,\b^2
-\alpha^*\left(\bc\right)^2
+\beta\,\b^4
+\beta^*\left(\bc\right)^4
\right]
\ket{n}
\ ,
\label{d2rhs}
\ee
where terms proportional to $\beta$ will be omitted from now on
since they do not affect the final power spectrum
[to order $\delta^2$, see Eq.~(\ref{Pk})], and
\be
\alpha_{\rm II}
=
\frac{3\,\alpha_{\rm I}}{2\,i\,a^3\mathcal{H}^3}
&\!\!=\!\!&
\frac{k^2}{a^5\,\mathcal{H}^4}
\left\{
2\,A\,B^*
\left[\left(\left|A\right|^2+\frac{1}{2}\right)
+\frac{i\,\mathcal{H}}{k/a}\right]
\right.
\nonumber
\\
&&
\phantom{\frac{k^2}{a^5\,\mathcal{H}^4}}
\ \
\left.
-\frac{2\,i\,\mathcal{H}}{k/a}
\left[\left(B^*\right)^2+A^2\right]
\right\}
\ .
\ee
\end{subequations}
Since
$\alpha_{\rm I}/\expec{\hat H_k}\sim a\,\mathcal{H}/k\lesssim |\delta|/k\ll 1$,
it appears that $\hat W_{\rm I}$ is actually negligible with respect
to $\hat H_k$.
In the regime~I, one can therefore choose suitable initial conditions
$\rho(t_{\rm i})$ and $\dot \rho(t_{\rm i})$ for which
\be
A(t_{\rm i})=0
\quad
{\rm and}
\quad
B(t_{\rm i})=1
\ ,
\label{initAB}
\ee
so that $\ket{n_E}=\ket{n}$ and $\ket{\chi_{\rm s}}=\ket{n}$
at $t=t_{\rm i}$.
This state will not change significantly until it enters
the regime~II and starts to be acted upon by $\hat W_{\rm II}$.
We can then assume a perturbative expansion of the form
\be
\ket{\chi_{\rm s}}=\ket{n_{\rm s}}\simeq
\ket{n}+\delta^2\,\ket{n^{(1)}}
=\left(\hat 1+\delta^2\,\hat R_n\right)\ket{n}
\ .
\label{firstorderans}
\ee
From this it then follows that
\be
\left(i\,\pt\hat R_n-\left[\hat H_k,\hat R_n\right]\right)\ket{n}
=\left[\alpha\,\hat b^2-\alpha^*\left(\hat b^\dagger\right)^2\right]
\ket{n}
\ ,
\label{Rneq}
\ee
which admits the solution
\begin{subequations} 
\be
\hat R_n \ket{n}=\left[r\, \hat b^2+r^*\left(\hat b^\dagger\right)^2\right]
\ket{n}
\ ,
\label{Rnsol}
\ee
provided $r$ satisfies the differential equation
\be
i\,\dot r+2\, \dot\Theta\, r=\alpha
\ .
\label{req}
\ee
\end{subequations}
Although $\hat R_n$ contains Hermitian parts, matter evolution is
unitary in general~\cite{bertoni} due to the form of Eq.~(\ref{keqtot0})
and in the case at hand one can in fact show that
$\langle\chi_{\rm s}|\chi_{\rm s}\rangle=1$ to order $\delta^2$
at all times for the states~(\ref{firstorderans}).
\subsection{Dispersion relations}
Two quantum states $\ket{\chi_{\rm s}}$ and $\ket{\bar \chi_{\rm s}}$
are physically indistinguishable when they share the same expectation values for
all the measurable observables $\hat X$,
\be 
\bra{\chi_{\rm s}}\,\hat X\, \ket{\chi_{\rm s}}
=\bra{\bar\chi_{\rm s}}\,\hat X\,\ket{\bar\chi_{\rm s}}
\ .
\ee
In this sense, since we are interested in observable quantities,
such as the power-spectrum, quadratic in $\hat \phi$ and
$\hat \pi_\phi$, we cannot distinguish between
$\ket{n_{\rm s}}$ and the state
\begin{subequations}
\be
\ket{\bar n_{\rm s}}=\left(\hat 1+i\,\delta^2 \hat H_n\right)
\ket{n}
\ ,
\label{eqevo}
\ee
where
\be
\hat H_n=i\,\frac{n^2+n+1}{2\,n+1}\,\left[
r\,\hat b^2-r^*\left(\hat b^\dagger\right)^2
\right]
\ ,
\ee
\end{subequations}
which evolves in an explicitly unitary way~\cite{toappear}.
\par
One can for instance consider the invariant vacuum,
$\ket{n_{\rm s}}=\ket{0}$ at $t=t_{\rm i}$, corresponding to an initial
state devoid of particles (of wavenumber $k$). 
The Hamiltonian associated with its effective evolution~(\ref{eqevo})
is given by 
\be
\hat H_n^{{\rm eff}}
&\!\!=\!\!&
\hat H_k
-\delta^2
\left[{\alpha\,\hat b^2+\alpha^*\left(\hat b^\dagger\right)^2}
\right] 
\nonumber
\\
&\!\!=\!\!&
\frac{1}{2}\left[
\frac{\hat \pi_k^2}{\mu}
+\mu\,\omega^2\,\hat\phi_k^2
\right]
+\gamma\left(\hat \phi_k\,\hat \pi_k
+\hat \pi_k\,\hat \phi_k\right)
\ ,
\label{Heff}
\ee
where $\omega$ is the effective frequency and,
for $\ell_{\rm p}^2/(a^3\mathcal{H})\ll 1$ and $a\,\mathcal{H}/k\ll 1$,
the effective mass is
\be
\mu\simeq
a^3\left[1-\frac{\ell_{\rm p}^2}{2\,a^3\,\mathcal{H}}\left(\frac{a\,\mathcal{H}}{k}\right)\right]
\ee
and 
\be
\gamma\simeq
-\frac{\ell_{\rm p}^2}{a^3}\left(\frac{k}{a\,\mathcal{H}}\right)
\ .
\ee
Apart from the squeezing term proportional to $\gamma$,
the perturbed dynamics can be described
by a modified dispersion relation~(\ref{mod_disp}).
In fact, Eq.~(\ref{pinney}) with the initial conditions~(\ref{initAB})
at $t_{\rm i}\to-\infty$ admits the solution
\be
\rho(t)=
\frac{1}{a\,k^{1/2}}\,\sqrt{1+\left(\frac{a\,\mathcal{H}}{k}\right)^2}
\ ,
\label{rho_app}
\ee
from which one obtains
\be
\omega
\simeq
\frac{k}{a}
\left[1+\frac{321\,\mathcal{H}^2\,\ell_{\rm p}^5}{64\,a^3}
\left(\frac{a}{k\,\ell_p}\right)^{3}\right]
\ ,
\label{modfreq}
\ee
valid for trans-Planckian modes with $a/k\,\ell_p\lesssim 1$.
The correction inside the square brackets is proportional to
the factor $(\ell_{\rm p}/a)^3$.
In a classical Robertson-Walker Universe there is no fundamental
length scale and one would not expect any dependence on
the numerical value of $a=a(t)$, since only its ratio at two different
times has a physical meaning.
However, in the quantum theory $\ell_{\rm p}$ plays the role
of the fundamental length which breaks scale invariance and
in terms of which all quantities with length dimension must be measured.
In fact, ``quantum gravitational fluctuations'' in the right hand side
of the matter Eq.~(\ref{kmateq}) are precisely proportional to
$(\ell_{\rm p}/a)^3$ and $(\mathcal{H}\,\ell_{\rm p})^2$.
\subsection{CMBR spectrum}
%
%
We can also compute the power spectrum
\be
\mathcal{P}_\phi(k)\equiv k^3\,\bra{\rm vac}\,\hat \phi_k^2\,\ket{\rm vac}
\ ,
\ee
and compare to the well-known case $\delta=0$.
The result will in general depend on the initial quantum state
we choose at $t=t_{\rm i}$ (see, {\em  e.g.}, Ref.~\cite{bozza}).
As we did for the dispersion relation, we choose the adiabatic invariant
vacuum
\be
\ket{\rm vac}=\ket{0_{\rm s}}
\ ,
\label{vacuum}
\ee
which initially coincides with the Bunch-Davies vacuum~\cite{birrell} and 
is known to give a flat, scale invariant spectrum for modes
well outside the horizon, $a/k\gg\mathcal{H}^{-1}$.
Adding the leading order corrections from Eq.~(\ref{d2rhs}) yields
\be
\mathcal{P}_\phi(k)
\simeq
k^3\left(\bra{0_{\rm s}}\,\hat\phi_k^2\,\ket{0_{\rm s}}
+2\,\delta^2\,\re\,\bra{0_{\rm s}}\,\hat\phi_k^2\,\hat R\,\ket{0_{\rm s}}
\right)
\ ,
\label{PSpert}
\ee
where 
\be
\bra{0_{\rm s}}\,\hat\phi_k^2\,\hat R\,\ket{0_{\rm s}}
=\frac{r^*}{k\,a^2}\,\left[\left(B^*\right)^2+A^2+2\,B^*A\right]
\ ,
\label{corrPS}
\ee
and $r=r(t)$ is a solution of Eq.~(\ref{req}).
We again employ Eq.~(\ref{rho_app}) for
trans-Planckian modes with $k\gtrsim a(t_{\rm i})/\ell_{\rm p}$
to evaluate the expression~(\ref{PSpert}) at 
$t=t_{\rm f}$ such that $a(t_{\rm f})\gg k/{\mathcal{H}}$,
and obtain
\be
\mathcal{P}_\phi(k)
\simeq
\frac{3}{8}\,\mathcal{H}^2\,
\left[1+\delta^2\left(u+\frac{v}{k^3}\right)\right]
\ ,
\label{Pk}
\ee
where $u$ is an integration constant related to the initial conditions
for $r=r(t)$ and $v\simeq 1.22$.
The zero order is precisely the expected flat spectrum, and 
$k$-dependent deviations are of order $\delta^2$, as we had
already anticipated from the general form of Eq.~(\ref{keqtot}).
\section{Conclusions}
\label{s_conc}
We have derived an effective modified dispersion relation~(\ref{modfreq})
for the trans-Planckian modes of a minimally coupled massless scalar
field in the de~Sitter space-time from the action principle~(\ref{Sp})
and the semiclassical treatment of the WDW equation~\cite{bv,acg}.
The effects induced by such modified dynamics on the power spectrum~(\ref{Pk}) 
during inflation appear as corrections of order $\delta^2=(\mathcal{H}\,\ell_{\rm p})^2$
and thus quite difficult to observe.
However, in order to obtain the expression in Eq.~(\ref{Pk}), we used the condition that
the initial state be the Bunch-Davies vacuum~(\ref{vacuum}).
Different results would follow from other choices, such as the one
considered in Ref.~\cite{ulf} which produces corrections of order
$\mathcal{H}\,\ell_{\rm p}$ and thus possibly
observable~\cite{ulf,act}, and the non-linearity of the matter
equation~(\ref{keqtot0}) may then lead to more non-trivial effects.
Our result~(\ref{Pk}) can thus be considered as a lower bound for the
size of quantum gravitational corrections.
\par
With regard to the two main issues mentioned in the
{\em Introduction\/}, we can therefore say that our work has shown:
a) that the semiclassical approach to the WDW equation
of Refs.~\cite{bv,acg} can consistently describe the evolution of
trans-Planckian matter modes starting from the very early stages and
b) that the initial quantum matter state can be more relevant than the
corrections produced during its subsequent semiclassical evolution
(see also Ref.~\cite{bozza} in this respect).
It appears then that initial conditions other than~(\ref{vacuum})
could only be justified by appealing to new physics beyond the
semiclassical level of quantum gravity.
Quite remarkably, earlier computations in the low energy limit of
some string and M-theories also yielded corrections of order
$(\mathcal{H}\,\ell_{\rm p})^2$
for the CMBR spectrum~\cite{kaloper}, thus in agreement with
our (lower bound) estimate.
\par
Let us end by mentioning that our present conclusions hold
for minimal coupling between matter and gravity
and may change in more general cases~\cite{toappear}. 
Moreover, the backreaction of the quantum fluctuations in the
gravitational equation~(\ref{Geq}), which was presently neglected,
is also worth analyzing.
\acknowledgments
We would like to thank A.O.~Barvinsky, A.Yu.~Kamenshchik, G.P.~Vacca
and G.~Venturi for comments and suggestions.
\end{document}